\def\dint{\mathop{\displaystyle \int}}

\documentclass[12pt,a4paper]{article}
\usepackage{amsmath}
\usepackage{amssymb}
\usepackage{graphicx}
\usepackage[usenames,dvipsnames]{xcolor}
\usepackage{mathpazo}
\usepackage{hyperref}
\usepackage{multimedia}
\usepackage{comment}
\usepackage[small,nohug,heads=vee]{diagrams}
\diagramstyle[labelstyle=\scriptstyle]
\definecolor{light-gray}{gray}{0.80}

\usepackage{fancyhdr}
\renewcommand{\maketitle}{
    \begin{center}
      \Large
        {\bf Field signature for apparently superluminal particle motion}
        \vskip .3 true cm
      \small
        Martin Land \\
        \vskip .3 true cm
        Department of Computer Science \\
        Hadassah College \\
        37 HaNevi'im Street, Jerusalem \\
email: martin@hadassah.ac.il
      \end{center}
      \vskip .5 true cm
}
\input{tcilatex}

\begin{document}
\title{}
\author{}
\maketitle

% \title{Field signature for apparently superluminal particle motion}
% \author{Martin Land}
% \address{Department of Computer Science, Hadassah College, 37 HaNeviim Street, Jerusalem}
% \ead{martin@hadassah.ac.il}
%
%
\begin{abstract}
In the context of Stueckelberg's covariant symplectic mechanics, Horwitz and
Aharonovich \cite{neutrinos} have proposed a simple mechanism by which a
particle traveling below light speed almost everywhere may exhibit a transit
time that suggests superluminal motion.  This mechanism, which requires precise
measurement of the particle velocity, involves a subtle perturbation affecting
the particle's recorded time coordinate caused by virtual pair processes.  The
Stueckelberg framework is particularly well suited to such problems, because it
permits pair creation/annihilation at the classical level.  In this paper, we study a
trajectory of the type proposed by Horwitz and Aharonovich, and derive the
Maxwell 4-vector potential associated with the motion.  We show that the
resulting fields carry a signature associated with the apparent superluminal
motion, providing an independent test for the mechanism that does not require
direct observation of the trajectory, except at the detector.     
\end{abstract}
\parindent=0cm \parskip=10pt

\section{Introduction}

The interpretation of antiparticles as negative energy particles propagating backward
in time was proposed by Stueckelberg \cite{Stueckelberg} in the context of his
covariant Hamiltonian theory of interacting spacetime events $x^\mu(\tau)$ 
evolving as functions of a Poincar\'{e}-invariant parameter $\tau$.  His goal was to
represent a particle/antiparticle process by a single worldline whose time coordinate
advances and retreats with respect to the laboratory clock as its instantaneous
energy changes sign under interaction with gauge fields.  In order to obtain a
well-posed electrodynamic theory, Sa'ad, Horwitz, and Arshansky \cite{saad}
generalized Stueckelberg's formalism by
introducing five $\tau$-dependent gauge fields.  The resulting pre-Maxwell theory
differs from conventional electrodynamics, but reduces to Maxwell theory at $\tau$
equilibrium.  An overview of pre-Maxwell electrodynamics can be found in \cite{pm_refs}.

In the context of Stueckelberg electrodynamics, Horwitz and Aharonovich
\cite{neutrinos} have proposed a simple mechanism by which a particle traveling
below light speed almost everywhere may exhibit a transit time that suggests
superluminal motion.  This mechanism is analogous to a mountain trail with a
switchback, on which a hiker may walk for 2 hours covering 12 km, but arrive at
a point only 8 km from the starting point, with an apparent average speed of 4
km/hr instead of 6 km/hr.  Exchanging time and space, an event may briefly
reverse its time direction without affecting its space motion.  In this case,
the particle will continue its progress in space coordinates from the source to
the detector, but the laboratory clock will report a difference between
time-of-start and time-of-arrival that is less than expected.  In this paper, we
study the Coulomb field expected from motion of this type and indicate a small
deviation from the field expected for continuous linear motion.  

\section{Pre-Maxwell electrodynamics}

In pre-Maxwell theory, interactions take place between events in
spacetime rather than between worldlines. Each event occurring at $\tau$
induces a current density in spacetime that disperses for large $\tau$
for free particles (and hence asymptotically).  Conversely, the current density
induces an electromagnetic field that acts on the events.  The {\em five} gauge
fields are $\tau$-dependent and invariant under local gauge
transformations that also depend on $\tau$.  The combined theory can be derived from
a unique scalar action and is integrable.  It is convenient to write
$x^5 = \tau$ and adopt the index convention
\begin{equation}
\lambda ,\mu ,\nu =0,1,2,3\qquad \qquad \qquad \alpha ,\beta ,\gamma =0,1,2,3,5
\ \ .
\end{equation}
The fields act on the events through the Lorentz force given by
\begin{equation}
M\;\ddot{x}^{\mu }=e_{0}\ f_{\;\;\;\alpha }^{\mu }(x,\tau )\,\dot{x}^{\alpha }
\qquad \qquad \qquad \frac{d}{d\tau }(-\frac{1}{2}M\dot{x}^{2})
=e_{0}\ f_{5\alpha }\dot{x} ^{\alpha }
\end{equation}
where the gauge invariant fields are formed from the potentials
\begin{equation}
f_{\alpha \beta }(x,\tau )=\partial _{\alpha }a_{\beta }(x,\tau )-\partial _{\beta
}a_{\alpha }(x,\tau )\;. 
\end{equation}
The fields are induced by event currents through the pre-Maxwell equations
\begin{equation}
\partial _{\beta }f^{\alpha \beta }(x,\tau )=\frac{e_{0}}{\lambda }j^{\alpha
}(x,\tau )=ej^{\alpha }(x,\tau ) \qquad \qquad \qquad \epsilon ^{\alpha \beta \gamma \delta \epsilon }\partial
_{\alpha }f_{\beta \gamma }(x,\tau )=0
\label{pmax}
\end{equation}
in which the parameter $\lambda$ acts as a coherence length.  The classical
current for an event $\xi^\mu(\tau)$ is
\begin{equation}
j^\alpha\left( x,\tau \right) =e \; \dot\xi^\alpha (\tau) \; \delta^4 \left(x-\xi(\tau) \right) 
\end{equation}
where $\xi^5 =1$.
Under the boundary conditions 
\begin{equation}
a^5\left( x,\tau \right)
\underset{\tau \rightarrow \pm \infty }{\xrightarrow{\hspace*{1.5cm}}} 0
\qquad \qquad j^{5}\left( x,\tau \right)\underset{\tau \rightarrow \pm \infty}
{\xrightarrow{\hspace*{1.5cm}}} 0 
\label{bc}
\end{equation}
the standard Maxwell theory is extracted as the equilibrium limit of (\ref{pmax})
by integration over the worldline
\begin{equation}
\left. 
\begin{array}{c}
\partial _{\beta }f^{\alpha \beta }\left( x,\tau \right) =ej^{\alpha }\left(
x,\tau \right) \\ 
\\ 
\partial _{\lbrack \alpha }f_{\beta \gamma ]}=0
\end{array}
\right\} \underset{\int d\tau }{\mbox{\quad}\xrightarrow{\hspace*{1cm}}
\mbox{\quad}}\left\{ 
\begin{array}{c}
\partial _{\nu }F^{\mu \nu }\left( x\right) =eJ^{\mu }\left( x\right) \\ 
\\ 
\partial _{\lbrack \mu }F_{\nu \rho ]}=0
\end{array}
\right.
\end{equation}
where 
\begin{equation}
F^{\mu \nu }(x)=d\tau \; f^{\mu \nu }\left( x,\tau
\right) \qquad A^{\mu }(x)=\int d\tau \; a^{\mu
}\left( x,\tau \right) \qquad  J^{\mu }(x)= d\tau \; j^{\mu
}\left( x,\tau \right). 
\label{c-cat-2}
\end{equation}
This integration has been called concatenation \cite{concat} and links the
event current $j^{\mu }\left( x,\tau \right) $ with the divergenceless 
Maxwell particle current
$ J^{\mu }(x)$ defined on the entire worldline.
The field equations lead to a wave equation
\begin{equation}
\partial _{\alpha }\partial ^{\alpha }a^{\beta }\left( x,\tau \right)
=\left( \partial _{\mu }\partial ^{\mu }-\;\partial _{\tau }^{2}\right)
a^{\beta }\left( x,\tau \right) =-ej^{\beta }\left( x,\tau \right)
\end{equation}
and the principal part Green's function \cite{green} is
\begin{equation}
G\left( x,\tau \right) =-{\frac{1}{{2\pi }}}\delta (x^{2})\delta \left( \tau
\right) -{\frac{1}{{2\pi ^{2}}}\frac{\partial }{{\partial {x^{2}}}}}\ {\frac{
{\theta (x^{2}-\tau ^{2})}}{\sqrt{{x^{2}-\tau ^{2}}}}}=D\left( x\right)
\delta (\tau )-G_{correlation}\left( x,\tau \right) .  \rule[-25pt]{0pt}{50pt}
\label{grn}
\end{equation}
The first term has support on the lightcone at instantaneous $\tau $, and
recovers the standard Maxwell Green's function under concatenation. The
second term has spacelike support (${x^{2}>\tau ^{2}\geq 0}$) and vanishes
under concatenation, so it may contribute to correlations but not to Maxwell
potentials.  Terms of this type have been studied in \cite{jigal}.

\section{Piecewise linear trajectories}

We define a particle trajectory as a set of piecewise linear event
segments
\begin{equation}
\xi \left( \tau \right) =\sum_{i=1}^n\xi_i\left( \tau \right) \Theta
_i\left( \tau \right) =\sum_{i=1}^n\left( u_i\tau +q_i\right)
\Theta_i\left( \tau \right)
\label{traj}
\end{equation}
where $u_i$ and $q_i$ are constant 4-vectors and 
$\Theta_i\left( \tau \right) $ is some combination of $\theta $
functions that defines the support of segment $i$.  The velocities
satisfy
\begin{equation}
u_i = \dot \xi_i (\tau) = \dfrac{d\xi_i}{d\tau} 
= \dfrac{1}{\sqrt{1-{\mathbf v}^2}} (1,{\mathbf v})
\qquad \qquad 
u^2_i=-1 \ \ .
\end{equation}
The trajectory is associated with an instantaneous event current
\begin{equation}
j\left( x,\tau \right) =e \sum_{i=1}^n\ u_i\Theta_i\left( \tau \right) \delta
^4\left( x-q_i-u_i\tau \right) 
\end{equation}
with delta function support. To handle the delta functions the instantaneous current
is spread along the worldline through 
\begin{eqnarray}
j_{\varphi }\left( x,\tau \right) &=&e \int ds~\varphi \left( \tau -s\right)
~j\left( x,s\right) \notag \\
&=&e \sum_{i=1}^n\int ds~\varphi \left( \tau -s\right)  u_i\Theta
_i\left( s\right)  \delta ^3\left( \mathbf{x}-\mathbf{q}_i-\mathbf{u}
_is\right) \delta \left( t-q_i^0-u_i^0s\right) \notag \\
&=&e \sum_{i=1}^n\int ds~\varphi \left( \tau -s\right) \dfrac{u_i\Theta
_i\left( s\right) }{\left\vert u_i^0\right\vert }\delta ^3\left( \mathbf{x}-
\mathbf{q}_i-\mathbf{u}_is\right) \delta \left( \dfrac{t-q_i^0}{
u_i^0}-s\right) \notag \\
&=&e \sum_{i=1}^n~\varphi \left( \tau -
\dfrac{t-q_i^0}{u_i^0}\right)  \dfrac{u_i}{\left\vert
u_i^0\right\vert }\Theta_i\left( \dfrac{t-q_i^0}{u_i^0}
\right)  
\delta ^3\left( \mathbf{x}-\mathbf{q}_i-\mathbf{u}_i\dfrac{t-q_i^0
}{u_i^0}\right) 
\label{curr}
\end{eqnarray}
where
\begin{equation}
\varphi \left( \tau \right) = \dfrac{1}{2\lambda}e^{-\left\vert\tau\right\vert /
2\lambda}
\qquad \qquad 
\int _\infty ^\infty d\tau \ \varphi (\tau) = 1 \ \ .
\end{equation}
The Maxwell current can be found by integrating this event current over $\tau$,
summing the event contributions along worldline and extracting the
equilibrium current
\begin{eqnarray}
J\left( x\right) &=&e \int d\tau ~\sum_{i=1}^n~\varphi \left( \tau -\dfrac{
t-q_i^0}{u_i^0}\right) \dfrac{u_i }{\left\vert u_i^0\right\vert } \Theta
_i\left( \dfrac{t-q_i^0}{u_i^0}\right) \delta ^3\left( \mathbf{x}-
\mathbf{q}_i-\mathbf{v}_i\left( t-q_i^0\right) \right) \notag\\
&=&e \sum_{i=1}^n~\dfrac{u_i }{\left\vert u_i^0\right\vert } \Theta_i\left( 
\dfrac{t-q_i^0}{u_i^0}\right) \delta ^3\left( \mathbf{x}-\mathbf{q}_i-
\mathbf{v}_i\left( t-q_i^0\right) \right) \ \ .
\end{eqnarray}
The Li\'{e}nard-Wiechert potential is found from the principal part Green's function
\begin{equation}
G(x-x^{\prime },\tau -\tau ^{\prime }) = -\delta \left[
\left( x-x^{\prime }\right) ^2\right] \delta \left( \tau -\tau ^{\prime
}\right)
\end{equation}
and the event current, so that
\begin{eqnarray}
a_{\varphi }(x,\tau ) &=& - {\dfrac{e}{{2\pi }}}\int d^4x^{\prime }d\tau ^{\prime }\ \delta \left[
\left( x-x^{\prime }\right) ^2\right] \delta \left( \tau -\tau ^{\prime
}\right) \ j_{\varphi }(x^{\prime },\tau ^{\prime }) \notag\\
&=& - {\dfrac{e}{{2\pi }}}\int d^4x^{\prime }\ \delta \left[ \left(
x-x^{\prime }\right) ^2\right] \ j_{\varphi }(x^{\prime },\tau ) \ \ .
\end{eqnarray}
Inserting (\ref{curr}) and performing the $x^\prime$ integration leads to
\begin{equation}
a_{\varphi }(x,\tau ) ={-\dfrac{e}{{2\pi }}}\sum_{i=1}^n\int d\tau ^{\prime }~\varphi (\tau
-\tau ^{\prime }) \ u_i\Theta_i\left( \tau ^{\prime }\right)
\delta \left[ \left( x-q_i-u_i\tau ^{\prime }\right)
^2\right] \ \ .
\end{equation}
Using the identity
\begin{equation}
\dint d\tau f\left( \tau \right) \delta \left[g\left( \tau \right) \right]
=\dfrac{f\left( s\right) }{\left\vert g^{\prime }\left( s\right)
\right\vert }
\end{equation}
we may perform the integral over $\tau^\prime$ to find
\begin{equation}
a_{\varphi }(x,\tau ) = -{\dfrac{e}{{4\pi }}}\sum_{i=1}^n\dfrac{
u_i\Theta_i\left( s_i\right)  \varphi (\tau -s_i)}{\left\vert
u_i\cdot \left( x-q_i-u_is_i\right) \right\vert } 
\end{equation}
where $s_i$ is determined by the fixed observation point $x$ and the 
\emph{a priori} trajectory $\xi_i\left( \tau \right) $ through 
\begin{equation}
\left( x-\xi_i\left( s_i\right) \right) ^2 = \left( x-q_i-u_is_i\right) ^2=0  \qquad \qquad 
x^0-q^0_i-u^0_is^0_i > 0 \ \ .
\end{equation}
Using $u_i^2=-1$ we find the retarded times as
\begin{equation}
s_i =-u_i\cdot \left( x-q_i\right) \mp \sqrt{\left( u_i\cdot \left(
x-q_i\right) \right) ^2+\left( x-q_i\right) ^2} \ \ .
\end{equation}
The choice of sign is associated with the retarded and advanced times of the
event evolution, and depends upon the direction of the
time evolution. Consider the ``static'' event for which $\mathbf{u}=0$ and $q=0$. For the
forward time trajectory $u = (1, \mathbf{0})$ 
\begin{equation}
s_\pm=t\mp \sqrt{\left( -t\right) ^2+\mathbf{x}^2-t^2}=t\mp R \ \ .
\end{equation}
At the observation point $\left( t,\mathbf{x}\right) $ the event observed 
must have been produced by the event current at a time earlier than $t$, and so we
choose the upper sign and retarded time $s_-$. The event produced at time $s_-$ is
located at $x=\left( t-R,\mathbf{ 0}\right) $ and as expected is at future lightlike
separation from the observation point. But for the ``static'' event evolving
backward in time with $u = (-1, \mathbf{0})$  
\begin{equation}
s_\pm=-t\mp \sqrt{\left( -t\right) ^2+\mathbf{x}^2-t^2}=-t\mp R
\end{equation}
and we must choose the lower sign. Of the two times $s_\pm$
it is clear that the event evolving toward negative times will
reach the point $s_+$ before reaching the point $s_-$ and so this is the
retarded time for that trajectory. At time $s_+$ the event is located at 
\begin{equation}
x=us_+=\left( -1,0\right) \left( -t+R\right) =\left( t-R,0\right) \ \ .
\end{equation}
Adding the lightlike
vector $\left( R,\mathbf{x}\right) $ takes this to the observation point $
\left( t,\mathbf{x}\right) $ as for the forward evolving event.
The choice of the upper sign can also be understood as choosing the $T$-reversed
picture of the forward time evolution, respecting the 
discrete Lorentz symmetry. We summarize
these cases as
\begin{equation}
s_i=-u_i\cdot \left( x-q_i\right) -\varepsilon \left( u_i^0\right) 
\sqrt{\left( u_i\cdot \left( x-q_i\right) \right) ^2+\left(
x-q_i\right) ^2}  
\label{s_R}
\end{equation}
and it follows that
\begin{equation}
\left\vert u_i\cdot \left( x-q_i-u_is_i\right) \right\vert =\sqrt{
\left( u_i\cdot \left( x-q_i\right) \right) ^2+\left( x-q_i\right)
^2}
\end{equation}
Finally, the Li\'{e}nard-Wiechert potential takes the form
\begin{equation}
a_{\varphi }(x,\tau )= - {\dfrac{e}{{4\pi }}}\sum_{i=1}^n\dfrac{
u_i\Theta_i\left( s_i\right)  \varphi (\tau -s_i)}{\sqrt{\left(
u_i\cdot \left( x-q_i\right) \right) ^2+\left( x-q_i\right) ^2}} 
\end{equation}
which under concatenation provides the Maxwell potential in the form
\begin{equation}
A(x)= - {\dfrac{e}{{4\pi }}}\sum_{i=1}^n\dfrac{u_i\Theta_i\left(
s_i\right)  }{\sqrt{\left( u_i\cdot \left( x-q_i\right) \right)
^2+\left( x-q_i\right) ^2}}  \ \ .
\end{equation}

\section{Field from pair annihilation}

We first consider a trajectory on which an event moves uniformly forward in
time until $\tau =0$ and then moves uniformly backward in time. In the
laboratory, this will we seen as pair annihilation. The event trajectory is
described by
\begin{equation}
\xi \left( \tau \right) =\left\{ 
\begin{array}{lll}
\left( u^0,\mathbf{u}\right) \tau  & ,
\rule[-024pt]{0cm}{024pt}
& \tau <0 \\ 
\left( -u^0,\mathbf{u}\right) \tau  & , & \tau >0
\end{array}
\right. 
\end{equation}
from which
\begin{equation}
u_1 =\left( u^0,\mathbf{u}\right) \qquad \mbox{\qquad}u_2=\left( -u^0,
\mathbf{u}\right) \mbox{\qquad}\qquad q_1=q_2=0 
\end{equation}
\begin{equation}
\Theta_1 =\left[ 1-\theta \left( \tau \right) \right] =\theta \left(
-\tau \right) \mbox{\qquad}\qquad\qquad  \Theta_2=\theta \left( \tau \right) 
\end{equation}
This trajectory is shown in Figure 1.
\begin{center}
\includegraphics[width=3.0in]{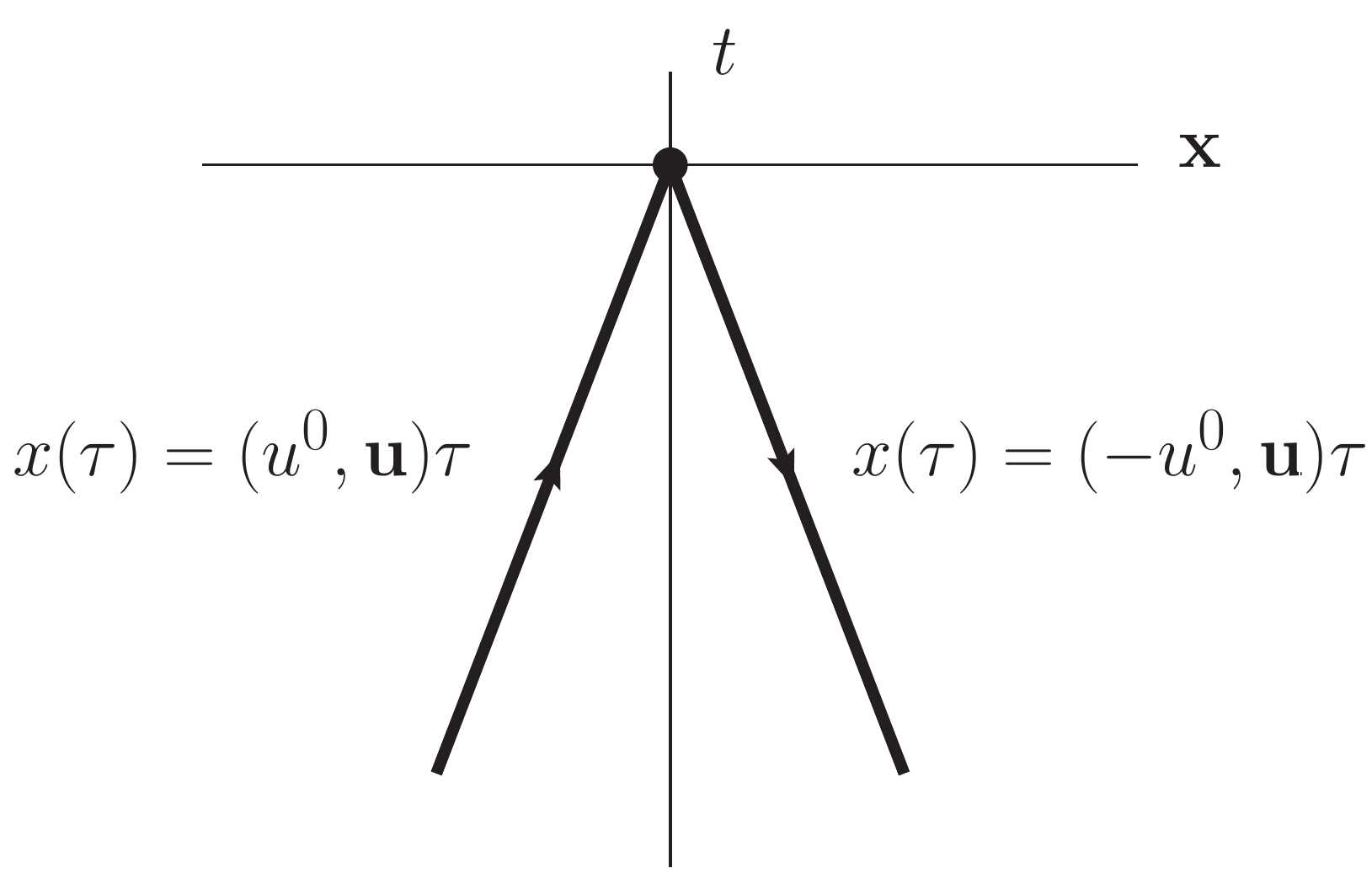}\\
\begin{tabular}{l}
{\footnotesize \ \textbf{Figure 1}: Pair annihilation event} $\hspace{7cm}$\\ 
\end{tabular}%

\end{center}
The event current takes the form
\begin{equation}
j\left( x,\tau \right)  =e \left( u^0,\mathbf{u}\right) \left[ 1-\theta \left( \tau \right) \right]
\delta ^4\left( x-\left( u^0,\mathbf{u}\right) \tau \right) +\left(
-u^0,\mathbf{u}\right) \theta \left( \tau \right) \delta ^4\left(
x-\left( -u^0,\mathbf{u}\right) \tau \right) 
\end{equation}
and the spread current is
\begin{eqnarray}
j_{\varphi }\left( x,\tau \right)  =&&\hspace{-18pt}e\sum_{i=1}^n\varphi \left( \tau -
\dfrac{t}{u_i^0}\right) \dfrac{u_i}{\left\vert u_i^0\right\vert }
\Theta_i\left( \dfrac{t}{u_i^0}\right) \delta ^3\left( \mathbf{x}-
\mathbf{u}_i\dfrac{t}{u_i^0}\right)  \\
=&&\hspace{-18pt}e\left[ \varphi (\tau -\dfrac{t}{u^0})\left( 1,\mathbf{v}\right)
\delta ^3\left( \mathbf{x}-\mathbf{v}t\right) \right. \notag \\
&&\qquad \qquad \qquad \qquad \left.
+\varphi (\tau +\dfrac{t}{u^0})
\left( -1,\mathbf{v}\right) \delta ^3\left( \mathbf{x}+\mathbf{v}
t\right) \right] \theta \left( -t\right) \ \ .
\end{eqnarray}
In this form, the first term describes a particle located at $\mathbf{x}=
\mathbf{v}t$ and the second term describes 
an antiparticle located at $\mathbf{x}=-\mathbf{v}t$, until
they mutually annihilate at $t=0$. Similarly, the concatenated current is
\begin{equation}
J\left( x,\tau \right) =e\left[ \left( 1,\mathbf{v}\right) ~\delta ^3\left( 
\mathbf{x}-\mathbf{v}t\right) +\left( -1,\mathbf{v}\right) ~\delta
^3\left( \mathbf{x}+\mathbf{v}t\right) \right] \theta \left( -t\right) 
\end{equation}
The retarded times are 
\begin{eqnarray}
s_1 &=&-\left( u^0,\mathbf{u}\right) \cdot x-\sqrt{\left( \left( u^0,
\mathbf{u}\right) \cdot x\right) ^2+x^2} \notag \\ 
&=&u^0t-\mathbf{u}\cdot \mathbf{x
}-R\sqrt{1+\left( \dfrac{t}{R}\right) ^2\left[ \left( \dfrac{\mathbf{u}
\cdot \mathbf{x}}{t}-u^0\right) ^2-1\right] } \\
s_2 &=&-\left( -u^0,\mathbf{u}\right) \cdot x+\sqrt{\left( \left( -u^0,
\mathbf{u}\right) \cdot x\right) ^2+x^2} \notag \\ 
&=&-u^0t-\mathbf{u}\cdot \mathbf{
x}+R\sqrt{1+\left( \dfrac{t}{R}\right) ^2\left[ \left( \dfrac{\mathbf{u}
\cdot \mathbf{x}}{t}+u^0\right) ^2-1\right] }
\end{eqnarray}
and the potential takes the form
\begin{eqnarray}
a^{\beta }(x,\tau )&=& - {\dfrac{e}{{4\pi }R}}\left[ \dfrac{\left( u^0,\mathbf{u
}\right) \varphi (\tau -s_1)\left[ 1-\theta \left( s_1\right) \right]   }{
\sqrt{1+\left( \dfrac{t}{R}\right) ^2\left[ \left( \dfrac{\mathbf{u}\cdot 
\mathbf{x}}{t}-u^0\right) ^2-1\right] }} \right. \notag \\ 
&&\qquad \qquad \qquad \qquad \qquad \left. +\dfrac{\left( -u^0,\mathbf{u}
\right) \varphi (\tau -s_2)\theta \left( s_2\right)   }{\sqrt{1+\left( 
\dfrac{t}{R}\right) ^2\left[ \left( \dfrac{\mathbf{u}\cdot \mathbf{x}}{t}
+u^0\right) ^2-1\right] }}\right] \ \ .
\end{eqnarray}
Under concatenation, the Maxwell potential is found to be
\begin{eqnarray}
A\left( x\right) &=& - {\dfrac{e}{{4\pi }R}}\left[ \dfrac{\left( u^0,\mathbf{u}
\right) \theta \left( -s_1\right)  }{\sqrt{1+\left( \dfrac{t}{R}\right) ^2
\left[ \left( \dfrac{\mathbf{u}\cdot \mathbf{x}}{t}-u^0\right) ^2-1
\right] }} \right. \notag \\
&&\qquad \qquad \qquad \qquad \qquad \left.
+\dfrac{\left( -u^0,\mathbf{u}\right) \theta \left( s_2\right)   
}{\sqrt{1+\left( \dfrac{t}{R}\right) ^2\left[ \left( \dfrac{\mathbf{u}
\cdot \mathbf{x}}{t}+u^0\right) ^2-1\right] }}\right] 
\label{coul}
\end{eqnarray}
We notice that the first term is nonzero for $s_1<0$ which holds in a region of
spacetime for which
\begin{eqnarray}
-\left( u^0,\mathbf{u}\right) \cdot x-\sqrt{\left( \left( u^0,\mathbf{u}
\right) \cdot x\right) ^2+x^2} &<&0 \\
\left[ \left( u^0,\mathbf{u}\right) \cdot x\right] ^2 &<&\left( \left(
u^0,\mathbf{u}\right) \cdot x\right) ^2+x^2 \\
0 &<&x^2
\end{eqnarray}
and the second term is nonzero for $s_2>0$ which requires
\begin{eqnarray}
-\left( -u^0,\mathbf{u}\right) \cdot x+\sqrt{\left( \left( -u^0,\mathbf{u
}\right) \cdot x\right) ^2+x^2} &>&0 \\
\left( \left( u^0,\mathbf{u}\right) \cdot x\right) ^2+x^2 &>&\left[
\left( u^0,\mathbf{u}\right) \cdot x\right] ^2 \\
x^2 &>&0 
\end{eqnarray}
so that
\begin{equation}
\theta \left( -s_1\right)  = \theta \left( s_2\right)  
= \theta \left( R -t\right) \ \ .
\end{equation}
The field (\ref{coul}) thus describes 
the combined field of two opposite charges that when observed from a fixed distance $R$
are seen to mutually annihilate at $t=R$. Put another way, the field is
observable only at spacelike separation from the annihilation event $\xi=0$.
At any point in the future of the annihilation event, there
is no observed field. We notice that if $\mathbf{u}\cdot \mathbf{x}=0$ the
component $A^0$ vanishes identically, describing a combined charge
distribution built from a pair of opposite charges viewed symmetrically from
the observation point $x$.

\section{Switchback in time}

We now consider an event evolving according to
\begin{equation}
\xi \left( \tau \right) =\left\{ 
\begin{array}{lll}
\left( u^0\tau ,\mathbf{u}\tau \right)  & ,
\rule[-024pt]{0cm}{024pt}
& \tau <0 \\ 
\left( -u^0\tau ,\mathbf{u}\tau \right)  & ,
\rule[-024pt]{0cm}{024pt}
& 0\leq \tau \leq T \\ 
\left( u^0\left( \tau -2T\right) ,\mathbf{u}\tau \right) =\left( u^0\tau
,\mathbf{u}\tau \right) +q_3 & , & \tau >T
\end{array}
\right. 
\end{equation}
where $q_3=\left( -2u^0T,\mathbf{0}\right) $. 
This trajectory is shown in Figure 2 --- its laboratory phenomenology is
discussed in section 6.
\begin{center}
\includegraphics[width=3.0in]{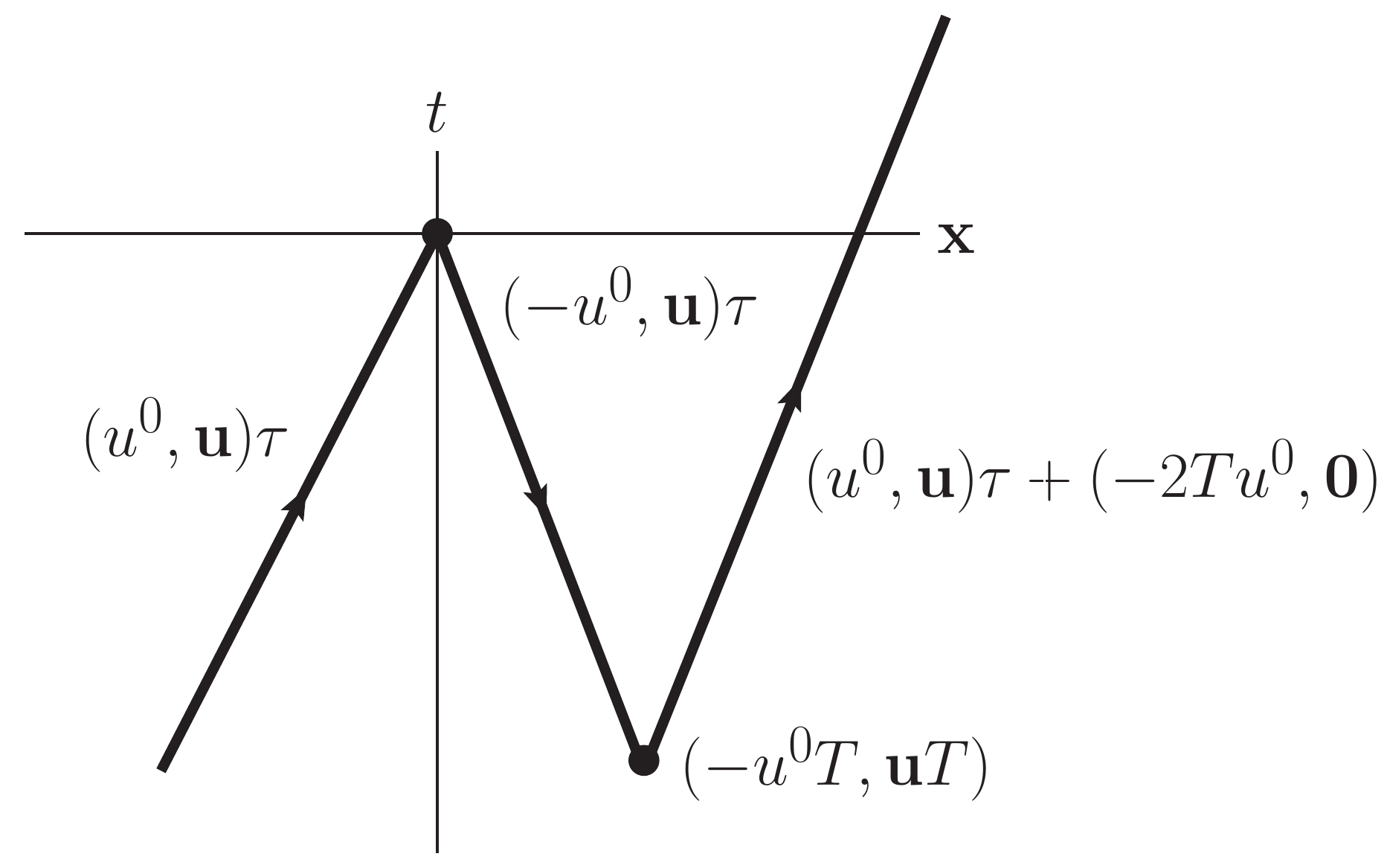}\\
\begin{tabular}{l}
{\footnotesize \ \textbf{Figure 2}: Switchback in time} $\hspace{7cm}$\\ 
\end{tabular}%

\end{center}
As described in the
Horwitz-Aharonovich model, this event proceeds moves linearly and continuously
in space, but experiences a brief switchback in its time coordinate.  For
example, if the particle is observed at $\tau = -\tau_0 < 0$ and observed again
at $\tau = \tau_0 > T$, the particle is seen to travel the distance 
$2\left\vert{\mathbf u} \right\vert \tau_0$
during an elapsed time $2u^0 \tau_0 -2T$ with speed   
\begin{equation}
v = \dfrac{\left\vert{\mathbf u} \right\vert}
{u^0\left(1-\dfrac{T}{\tau_0}\right)} < \dfrac{\left\vert{\mathbf u} \right\vert}{u^0} \ \ .
\end{equation}
The four-velocity for this trajectory is
\begin{equation}
\dot{\xi}\left( \tau \right) =\left\{ 
\begin{array}{lll}
\left( u^0,\mathbf{u}\right)  & ,
\rule[-024pt]{0cm}{024pt}
& \tau <0 \\ 
\left( -u^0,\mathbf{u}\right)  & ,
\rule[-024pt]{0cm}{024pt}
& 0\leq \tau \leq T \\ 
\left( u^0,\mathbf{u}\right)  & , & \tau >T
\end{array}
\right. 
\end{equation}
and current is
\begin{equation}
j\left( x,\tau \right) =e \left\{ 
\begin{array}{lll}
\left( u^0,\mathbf{u}\right) \delta ^3\left( \mathbf{x}-\mathbf{u}\tau
\right) \delta \left( t-u^0\tau \right)  & ,
\rule[-024pt]{0cm}{024pt}
& \tau <0 \\ 
\left( -u^0,\mathbf{u}\right) \delta ^3\left( \mathbf{x}-\mathbf{u}\tau
\right) \delta \left( t+u^0\tau \right)  & ,
\rule[-024pt]{0cm}{024pt}
& 0\leq \tau \leq T \\ 
\left( u^0,\mathbf{u}\right) \delta ^3\left( \mathbf{x}-\mathbf{u}\tau
\right) \delta \left( t-u^0\left( \tau -2T\right) \right)  & , & \tau >T
\end{array}
\right. \ \ .
\end{equation}
which can be put into the form (\ref{traj}) as 
\begin{eqnarray}
j\left( x,\tau \right)  &=&e \left( u^0,\mathbf{u}\right) \delta ^3\left( 
\mathbf{x}-\mathbf{u}\tau \right) \delta \left( t-u^0\tau \right) \theta
\left( -\tau \right) \notag \\
&&+e \left( -u^0,\mathbf{u}\right) \delta ^3\left( 
\mathbf{x}-\mathbf{u}\tau \right) \delta \left( t+u^0\tau \right) \left[
\theta \left( \tau \right) -\theta \left( \tau -T\right) \right]  \notag\\
&&+e \left( u^0,\mathbf{u}\right) \delta ^3\left( \mathbf{x}-\mathbf{u}
\tau \right) \delta \left( t-u^0\left( \tau -2T\right) \right) \theta
\left( \tau -T\right) \ \ .
\end{eqnarray}
Using (\ref{s_R}) we find the retarded times
\begin{eqnarray}
s_1 &&\hspace{-18pt}=u^0t-\mathbf{u}\cdot \mathbf{x}-\sqrt{\left[ \mathbf{u}\cdot \mathbf{x}
-u^0t\right] ^2+R^2-t^2} \label{s1} \\
s_2 &&\hspace{-18pt}=-u^0t-\mathbf{u}\cdot \mathbf{x}+\sqrt{\left[ \mathbf{u}\cdot \mathbf{x}
+u^0t\right] ^2+R^2-t^2} \label{s2} \\
s_3 &&\hspace{-18pt}=u^0\left( t+2u^0T\right) -\mathbf{u}\cdot \mathbf{x}-\sqrt{\left[ 
\mathbf{u}\cdot \mathbf{x}-u^0\left( t+2u^0T\right) \right]
^2+R^2-\left( t+2u^0T\right) ^2} \label{s3}
\end{eqnarray}
so that the potential is
\begin{eqnarray}
a_{\varphi }(x,\tau ) &=& - {\dfrac{e}{{4\pi R}}}\left\{ \dfrac{\left( u^0,\mathbf{u}\right)
\varphi (\tau -s_1)\theta \left( -s_1\right)  }{\sqrt{1+\left( \dfrac{t}{R
}\right) ^2\left[ \left( \dfrac{\mathbf{u}\cdot \mathbf{x}}{t}
-u^0\right) ^2-1\right] }}\right.  {\rule[-052pt]{0cm}{052pt}} \notag \\
&&\hspace{2.0cm}+\dfrac{\left( -u^0,\mathbf{u}\right) \varphi (\tau -s_2)\left[ \theta
\left( s_2\right) -\theta \left( s_2-T\right) \right]  }{\sqrt{1+\left( 
\dfrac{t}{R}\right) ^2\left[ \left( \dfrac{\mathbf{u}\cdot \mathbf{x}}{t}
+u^0\right) ^2-1\right] }} {\rule[-024pt]{0cm}{024pt}} \notag \\
&&\hspace{1cm}\left. +\dfrac{\left( u^0,\mathbf{u}\right) \varphi (\tau -s_3)\theta
\left( s_3-T\right)  }{\sqrt{1+\left( \dfrac{t+2u^0T}{R}\right) ^2\left[
\left( \dfrac{\mathbf{u}\cdot \mathbf{x}}{\left( t+2u^0T\right) }
-u^0\right) ^2-1\right] }}\right\}  \ \ .
\end{eqnarray}
Under concatenation, the Maxwell potential becomes
\begin{eqnarray}
A\left( x\right)  &=& - {\dfrac{e}{{4\pi R}}}\left\{ \dfrac{\left( u^0,
\mathbf{u}\right) \theta \left( -s_1\right)  }{\sqrt{1+\left( \dfrac{t}{R}
\right) ^2\left[ \left( \dfrac{\mathbf{u}\cdot \mathbf{x}}{t}-u^0\right)
^2-1\right] }} \right.   {\rule[-052pt]{0cm}{052pt}} \notag \\
&&\hspace{3cm}+\dfrac{\left( -u^0,\mathbf{u}\right) \left[ \theta \left(
s_2\right) -\theta \left( s_2-T\right) \right]  }{\sqrt{1+\left( \dfrac{t
}{R}\right) ^2\left[ \left( \dfrac{\mathbf{u}\cdot \mathbf{x}}{t}
+u^0\right) ^2-1\right] }} {\rule[-024pt]{0cm}{024pt}} \notag \\
&&\hspace{1cm}\left. +\dfrac{\left( u^0,\mathbf{u}\right) \theta \left( s_3-T\right)  
}{\sqrt{1+\left( \dfrac{t+2u^0T}{R}\right) ^2\left[ \left( \dfrac{
\mathbf{u}\cdot \mathbf{x}}{\left( t+2u^0T\right) }-u^0\right) ^2-1
\right] }}\right\} \ \ .
\end{eqnarray}
As seen in the pair annihilation problem, 
the functions $\theta \left( -s_1\right) $ 
and $\theta \left( s_2\right) $ become $\theta\left(R-t\right)$ and we must
similarly convert $\theta \left( s_2-T\right) $ and $\theta \left( s_3-T\right)
$ to functions on spacetime. 
Using (\ref{s2}) the condition $s_2>T$ becomes
\begin{equation}
-u^0t-\mathbf{u}\cdot \mathbf{x}+\sqrt{\left[ \mathbf{u}\cdot 
\mathbf{x}+u^0t\right] ^2+R^2-t^2}>T
\end{equation}
which can be put into the form
\begin{equation}
t^2+2Tu^0t+T^2+2T\mathbf{u}\cdot \mathbf{x}-R^2 <0 \ \ .
\end{equation}
Since $u_i^2=-1$ this can be rewritten as
\begin{equation}
-\left( u_2T-x\right) ^2 <0 \ \ ,
\label{s2-gen}
\end{equation}
which is equivalent to the requirement that $u_2T-x$ be spacelike.  For this
term to be nonzero the observation
point $x$ must not be in the future of the event $u_2T$, the second turn-around
point at which the trajectory
returns to forward time evolution.  Using (\ref{s3}) the condition $s_3>T$ is
expressed as 
\begin{equation}
u^0\left( t+2u^0T\right) -\mathbf{u}\cdot \mathbf{x}-\sqrt{\left[
\mathbf{u}\cdot \mathbf{x}-u^0\left( t+2u^0T\right) \right]
^2+R^2-\left( t+2u^0T\right) ^2}>T
\end{equation}
equivalent to the requirement
\begin{equation}
-\left[ \left( u_3T-q_3\right) -x\right] ^2 >0
\end{equation}
where again $q_3=\left( -2u^0T,\mathbf{0}\right) $.  
Since 
\begin{equation}
u_3T-q_3=u_2T
\end{equation}
this requires that the observation point
be in the future of the second turn-around point.  It is convenient to write
\begin{equation}
\theta\left(s_3 - T\right) = \theta\left(-\left( x-u_2T\right)^2\right) 
= 1 - \theta\left(\left( x-u_2T\right)^2\right)
\end{equation}
which splits the potential into 3 pieces
\begin{equation}
A(x) =  A_1 (x) + A_2 (x) + A_3 (x) 
\label{A}
\end{equation}
where 
\begin{equation}
A_1 =-\dfrac{e}{4\pi R} \dfrac{\left( u^0,\mathbf{u}\right) \theta(t+u^0T)}{\sqrt{1+\left( \dfrac{
t+2u^0T}{R}\right) ^2\left[ \left( \dfrac{\mathbf{u}\cdot \mathbf{x}}{
\left( t+2u^0T\right) }-u^0\right) ^2-1\right] }}
\label{A-1}
\end{equation}
appears most like the usual Li\'{e}nard-Wiechert potential with 
support on the future of the second turn-around point, and 
\begin{eqnarray}
A_2 &=&-\dfrac{e}{4\pi R} \left[ \dfrac{\left( u^0,
\mathbf{u}\right) }{\sqrt{1+\left( \dfrac{t}{R}\right) ^2\left[ \left( 
\dfrac{\mathbf{u}\cdot \mathbf{x}}{t}-u^0\right) ^2-1\right] }} \right. \notag \\
&&\hspace{3cm}\left. +\dfrac{
\left( -u^0,\mathbf{u}\right) }{\sqrt{1+\left( \dfrac{t}{R}\right) ^2
\left[ \left( \dfrac{\mathbf{u}\cdot \mathbf{x}}{t}+u^0\right) ^2-1
\right] }}\right] \theta \left( R-t\right) 
\label{A-2}\\ 
A_3 &=&\dfrac{e}{4\pi R} \left[ \dfrac{\left( u^0,\mathbf{u}\right) }{\sqrt{1+\left( \dfrac{
t+2u^0T}{R}\right) ^2\left[ \left( \dfrac{\mathbf{u}\cdot \mathbf{x}}{
\left( t+2u^0T\right) }-u^0\right) ^2-1\right] }} \right. \notag \\
&&\hspace{2cm}\left. +\dfrac{\left( -u^0,
\mathbf{u}\right) }{\sqrt{1+\left( \dfrac{t}{R}\right) ^2\left[ \left( 
\dfrac{\mathbf{u}\cdot \mathbf{x}}{t}+u^0\right) ^2-1\right] }}\right]
\theta \left(  R-\left(t+u_2T\right) \right)
\label{A-3}
\end{eqnarray}
have support in the present (spacelike separation) of the ``pull-back'' in time. 

\section{Discussion}

In the previous section we derived the Maxwell potential associated with a
particle trajectory that includes a ``pull-back'' in time.  Under the
Stueckelberg-Feynman interpretation, the resulting zig-zag in the time
coordinate will be observed in the laboratory as the following sequence: 
\begin{enumerate}
	\item Particle-1 moves on a linear trajectory with energy $E = mu^0$ and
	velocity $\mathbf{u} / u^0$,
	\item At clock time $t = -u^0T$, particle-1 reaches ${\mathbf x} = 
	-\mathbf{u}T$, while a pair creation process at
	${\mathbf x}=\mathbf{u}T$ produces particle-2 and an antiparticle,
	\item At clock time $t=0$, particle-1 and the antiparticle mutually
	annihilate,
	\item Particle-2 continues on a linear trajectory congruent with the original path of particle-1.
\end{enumerate}
The observation of such a process and the measurement of the time ``pull-back''
would require many detectors along the particle trajectory.  Given the good
fortune to have placed detectors in (spacelike) proximity to the pair processes, one may
also observe the unusual contributions (\ref{A-2}) and (\ref{A-3})
to the Maxwell potential --- and hence the
electromagnetic fields.   

However, in addition to these signatures, the duration $T$ of the ``pull-back''
in time will be embedded, through equation (\ref{A-1}), in the form of the
electromagnetic fields at any point along the trajectory, specifically 
when the particle arrives at the detector.  If the line of observation is
transverse to the particle velocity, so that $\mathbf{u}\cdot \mathbf{x} = 0$,
then at high energy, with $\left(u^0\right)^2 \gg 1$, one may have 
$t \simeq 2u^0T$ causing the potential
\begin{equation}
A_1 \simeq -\dfrac{e}{4\pi R} \dfrac{\left( 1,\mathbf{v}\right) }{\sqrt{1+\left( \dfrac{
t+2u^0T}{R}\right) ^2 }}
\label{final}
\end{equation}
to deviate measurably from the expected value (found from equation (\ref{final}) with $T=0$).

%\section*{References}

%

%

%
\end{document}